# Interactions of nanorod particles in the strong coupling regime

Cheng-ping Huang[1*], Xiao-gang Yin[2], Ling-bao Kong[3], and Yong-yuan Zhu[2*]

[1]*Department of Applied Physics, Nanjing University of Technology*
*Nanjing 210009, P.R. China*

[2]*National Laboratory of Solid State Microstructures, Nanjing University*
*Nanjing 210093, P.R. China*

[3]*School of Science, Beijing University of Chemical Technology,*
*Beijing 100029, P.R. China*

Abstract

The plasmon coupling in a nanorod dimer obeys the exponential size dependence according to the Universal Plasmon Ruler Equation. However, it was shown recently that such a model does not hold at short nanorod distance (*Nano Lett.* **2009**, 9, 1651). Here we study the nanorod coupling in various cases, including nanorod dimer with the asymmetrical lengths and symmetrical dimer with the varying gap width. The asymmetrical nanorod dimer causes two plasmon modes: one is the attractive lower-energy mode and the other the repulsive high-energy mode. Using a simple coupled LC-resonator model, the position of dimer resonance has been determined analytically. Moreover, we found that the plasmon coupling of symmetrical cylindrical (or rectangular) nanorod dimer is governed uniquely by gap width scaled for the (effective) rod radius rather than for the rod length. A new Plasmon Ruler Equation without using the fitting parameters has been proposed, which agrees well with the FDTD calculations. The method has also been extended to study the plasmonic wave-guiding in a linear chain of gold nanorod particles. A field decay length up to 2700nm with the lateral mode size about 50nm ($\sim \lambda/28$) has been suggested.

* Email: cphuang@njut.edu.cn, yyzhu@nju.edu.cn

# I. Introduction

It is well known that the metallic nanoparticles can support localized plasmon resonance when illuminated with light of certain frequencies, where the confined conduction electrons oscillate resonantly under the action of light electric field. Besides the greatly enhanced light scattering and absorption, the plasmon resonance of metal particle is accompanied by a strong field enhancement near the particle surface. Such effect can be employed to concentrate light into deep-subwavelength volumes or improve the absorption of light in photovoltaic devices [1, 2]. Moreover, the particle plasmon resonance is dramatically dependent on the particle composition, shape, size, and local dielectric environment. This enables us to tune the plasmon resonance by using the composite core/shell structures or nonspherical nanoparticles such as the gold nanorods [3, 4], and to construct plasmonic nanosensors for sensing the refractive-index changes or detecting the bounded organic molecules [5]. It is worthy of noticing that the metallic particles can also serve as the fundamental building blocks for the design of various plasmonic microstructures, including one-dimensional plasmonic waveguides [6, 7], two-dimensional plasmonic band-gap materials [8, 9], and three-dimensional plasmonic polaritonic crystals [10, 11] etc.

On the other hand, the plasmonic coupling between a pair of metallic nanoparticles (called a dimer), e.g., nanospheres [12-15], nanodisks [16-19], nanoshells [20, 21], and nanorods [22-25], has also attracted much research interest. These particle dimers or plasmonic "molecules" can be viewed as a basic system for studying the plasmonic coupling effect and may be used to simulate the interactions between individual molecules. When two metallic particles approach each other, their plasmonic near-fields will overlap and couple strongly, giving rise to a distance-dependent wavelength shift of the plasmon mode. The effect can be understood using the plasmon hybridization method [26-28], where the coupled mode is treated as bonding and antibonding combinations of the individual particle plasmon modes. The distance dependence of plasmon coupling provides a useful tool or Plasmon Ruler for the measurement of distance between two metallic particles in biological systems [29].

Moreover, even stronger field enhancement can be achieved near the junction of particle dimer, which is critical for the trapping of single molecules [30], surface-enhanced Raman scattering [31], and nonlinear-optical frequency mixing [32] etc. And in recent years, the plasmonic optical antennas composed of a metallic nanorod dimer have also been investigated [33-35].

For light polarization along the interparticle axis, the dimer plasmon mode will be redshifted significantly with the decrease of particle separations. It has been suggested that, instead of the cubic size decay predicted by the simple point-dipole model, the size scaling of this plasmon coupling can be well described with a universal exponential distance dependence (or Plasmon Ruler Equation) [29, 36-40]

$$\Delta\lambda/\lambda_0 = \sigma \exp(-s/\tau l). \qquad (1)$$

Here, $\lambda_0$ is the plasmon resonance wavelength of an isolated metal particle, $\Delta\lambda$ is the wavelength shift of dimer with respect to the single particle, $l$ and $s$ is, respectively, the size and separation of particles, and $\sigma$, $\tau$ are two fitting parameters. This interesting scaling behavior, which was found to be valid for various particle shape and size [36-40], provides us a simple and convenient method for the study and application of the plasmon coupling effect. Nonetheless, the accuracy of Plasmon Ruler Equation at close approaches of metal particles has been challenged most recently. In the work by Funston et al. [25], the coupling between a pair of gold nanorods has been studied, where the interparticle distance is down to about 2nm. Their results show that for very small particle separations (gap width scaled by the rod length $s/l$<0.1) the exponential size scaling is not correct. This hampers the application of Plasmon Ruler Equation in the strong coupling regime.

In this paper, we aim at the plasmon coupling of gold nanorod particles within a dimer at close approaches. Several cases have been considered, including cylindrical nanorods with the asymmetrical lengths, where the plasmon modes are split with the fields concentrated in either nanorod, and symmetrical cylindrical nanorods with the decreasing gap width, where the plasmon mode is redshifted and the near-field is increasingly boosted. The rectangular nanorod dimer with the varying gap width has also been studied, suggesting the similar coupling behavior as that of the cylindrical one. To understand the nanorod coupling, a coupled LC-resonator model has been employed, which captures well the observed effect. Our results suggest that the plasmon coupling of cylindrical or rectangular nanorod dimer is uniquely determined by gap width scaled for the (effective) rod radius (s/r) rather than for the rod length (s/$l$) as taken previously [25, 38, 39]. A new Plasmon Ruler Equation for the symmetrical nanorod dimer has been proposed, which shows good agreement with the numerical calculations. Compared with Eq. (1), a remarkable feature of our result is that no fitting parameters are used. The method has also been extended to study the plasmonic wave-guiding in a linear chain of gold nanorod particles. A field decay length up to 2700nm with the lateral mode size about 50nm ($\sim \lambda/28$) has been suggested. These results may provide new opportunity for the understanding and application of the plasmon coupling effect.

The paper is structured as follows. In Sec. II, the structure investigated and the numerical simulation method are briefly described. Section III focuses on the plasmon coupling of gold nanorods within an asymmetrical cylindrical dimer. To understand, a coupled LC-resonator model is presented. In Sec. IV, we study the plasmon coupling of a symmetrical cylindrical dimer with the varying gap width. A new Plasmon Ruler Equation is proposed. The symmetrical rectangular nanorod dimer will be considered in Sec. V. And in Sec. VI, the plasmonic coupling in a linear array of gold nanorods is investigated, which provides an efficient channel for the energy transfer. A short summary is given in Sec. VII.

## II. Structure and simulation

In the nanorod dimer we are interested in, the two gold nanorods are aligned end-to-end, where a small gap is present which influences significantly the near- and far-field optical properties of the system. The nanorod dimer is embedded in a dielectric host medium with the permittivity of $\varepsilon_d$, and the incident light propagates with the electric field along the rod axis, thus exciting the longitudinal plasmon resonance. Here, the length of gold nanorods is typically in the deep subwavelength scale, and the transverse dimensions (circular or rectangular) of two nanorod particles are identical and smaller than twice the skin depth of gold (the fields within the particle can be taken to be homogeneous). And, the periodic arranging of gold nanorods along the rod axis (in the host medium) forms a plasmonic waveguide.

To explore the plasmon coupling effect of nanorods and verify the validity of the theoretical results, the commercial software package FDTD Solutions 6.5 (Lumerical Solutions, Inc., Canada) has been used in the following study. This software is based on the finite-difference time-domain method. And recently, it has been employed to determine the plasmonic coupling of silver sphere dimers [14]. In our FDTD simulations, the non-uniform mesh and smaller mesh size were used in order to get the reliable results. Near the gap region (including the gap width $s$ and 5nm length extending into each nanorod), the mesh size was defined as 0.5 nm (when $s>3nm$) or 0.25nm (when $1 \le s \le 3nm$). Whereas in the remaining region (within and outside the nanorods), the mesh size was set as 1.0 nm. For the single nanorods and nanorod dimers, a 3D total field scattered field (TFSF) source was used, which simulates an incident plane wave with the wavevector (polarization) perpendicular (parallel) to the rod axis. The total power of the scattered field was detected and normalized to that of the source. And, for the linear array of gold nanorods (end-to-end), an electric dipole source (parallel to the rod axis) was placed at a distance of 50nm away from the first rod. The dipole source excites the longitudinal plasmon wave, which propagates along the linear chain and can be detected by the monitor. Here, for all the simulations, the gold is modeled with a lossy Drude dispersion $\varepsilon(\omega) = \varepsilon_\infty - \omega_p^2 / \omega(\omega + i\gamma)$, where $\varepsilon_\infty = 7$, $\omega_p = 1.37 \times 10^{16}$ rad/s, and $\gamma = 1.0 \times 10^{14}$ rad/s, and the permittivity of host medium is set as $\varepsilon_d = 2.25$.

### III. Cylindrical nanorods of asymmetrical lengths

Recently, the plasmonic heterodimers due to structure or composition asymmetry have received much attention [41-44]. Here, we consider the case that the two metal particles are the cylindrical nanorods, which have the same radius $r$ but the asymmetrical rod lengths $l_1$ and $l_2$, respectively [see Fig. 1(b)]. In this case, the plasmon modes of individual particles are nondegenerate, and at close approaches they will couple strongly to each other. Without loss of generality, Fig. 1(a) presents the simulated far-field scattering spectra of the gold nanorod dimer, where the rod radius and gap width are fixed as r=s=15nm and the rod lengths are set as $(l_1, l_2)$=(120, 120), (120, 150), and (120, 180) nm, respectively. Compared with the isolated gold nanorods which have, respectively, the lengths 120, 150, 180nm and the resonance wavelength 1023, 1181, and 1335nm [see the inset of Fig.1(a)], the spectral positions of nanorod dimer modes have been significantly modified due to the plasmon coupling. For the nanorod dimer with $(l_1, l_2)$=(120, 120) nm, a single resonance peak is found in the spectrum (the black open circles), which is redshifted from 1023nm of isolated nanorod to a larger wavelength 1158nm. The axial electric-field distribution for the resonance mode is shown in Fig. 1(c). One can see that the two symmetrical nanorods are equally excited and that the middle dielectric gap exhibits a strong enhancement of axial electric field. However, for the asymmetrical nanorod dimer with $(l_1, l_2)$=(120, 150) nm, two resonance peaks can be observed (the green open squares), one locates at 980nm and the other at 1275nm. And with the increase of dimer asymmetry [$(l_1, l_2)$=(120, 180) nm], the two resonance peaks are redshifted to 1006nm and 1412nm, respectively (see the red solid circles). Figure 1(d) and 1(e) plot the axial electric-field distribution associated with the short (1006nm) and long (1412nm) wavelength resonance of the nanorod dimer. The results suggest that both nanorods can be excited at the two resonance wavelengths. Nonetheless, at the wavelength 1006nm the 120nm-long nanorod is strongly excited, and the other nanorod is more strongly excited at the longer wavelength 1412nm instead.

To study the above effect, we have employed here a simple analytical model. Recently, the concept of circuit elements at optical frequencies has been proposed by Engheta et al. [45-47], using the plasmonic and nonplasmonic nanoparticles. With the use of displacement current, they calculated the impedance of nanoelements and showed that a nanoparticle will act as a nanocapacitor or a nanoinductor, depending on the sign of the particle permittivity. And, in a recent publication, we proposed in a different way that a single gold nanorod can be treated as an LC circuit and that the plasmon resonance of nanorod can thus be determined [48]. In our approach, we emphasized the role of conduction current in a metal, which constitutes at optical frequencies the major part of the total current [in the metal, $\partial D/\partial t$ represents actually the total current, where $D = \varepsilon_0 \varepsilon(\omega) E$ and $\varepsilon(\omega)$ is the metal permittivity]. Under the action of light electric field, the conduction current is generated in the nanorod and positive and negative charges $\pm q$ will accumulate on the opposite sides. Consequently, the two oppositely charged end faces (circular disks) of the nanorod will function as a circular capacitor. The inductance includes two parts [48], one is the self-inductance $L$ associated with the magnetic field energy and the other is a formal inductance $L_0$ relating to the electronic inertia or kinetic energy (in other words, the electronic inertia acts as an inductance; note that $L_0$ corresponds to that reported in Ref. [45-47]). Such an LC resonator supports a localized resonance, where the confined electrons behave as the forced harmonic oscillators characterized by an effective restoring force and an increased effective mass (the self-inductance acts as the electronic inertia, instead) [11]. In addition to the plasmon resonance wavelength, this model also enables us to calculate the dipole moment and extinction spectrum of the nanorod (especially when the rod length $l << \lambda$). Correspondingly, the plasmon coupling in a nanorod dimer may be treated by the coupling of two LC resonators.

We assume that the electric charges carried by the two asymmetrical nanorods are $\pm q_1$ and $\pm q_2$, respectively. Due to the coupling, the potential difference of one nanorod (between the centers of two end faces) will benefit not only from the charges of itself but also from that of the other nanorod. Especially, at the close approaches, the coupling of nanorods is mediated mainly by the terminal surface charges close to the dielectric gap. In this case, the potential difference of nanorods can be expressed as $\Delta u_1 = q_1/C_{11} - q_2/C_{12}$ and $\Delta u_2 = q_2/C_{22} - q_1/C_{21}$, where $C_{11} = C_{22} = \alpha\pi\varepsilon_0\varepsilon_d r$ and $C_{12} = C_{21} = 2\alpha\pi\varepsilon_0\varepsilon_d r^2/(\sqrt{r^2+s^2}-s)$ (here $s$ is the gap width and $\alpha$ is the fraction of charges distributed on the end faces). Since the wavevector of incident light is perpendicular to the nanorod axis and the wavelength in the host medium (~1000nm) is much larger than the dimer width (~30nm), the quasi-static approximation is applicable. Therefore, the circuit equation $El - \Delta u = I(R - i\omega L)$ can be applied, respectively, to each subwavelength gold nanorod ($E$ is the light electric field, $R = R_0 - i\omega L_0$ is the ac resistance or impedance [48]). Noticing that each nanorod (carrying opposite charges on the opposite sides) can be regarded as an electric dipole with its dipole moment $p \approx ql$, we have

$$p_1 = \frac{\beta_1(\omega_2^2 - \omega^2 - i\eta_2\omega) + \alpha_1\beta_2}{(\omega_1^2 - \omega^2 - i\eta_1\omega)(\omega_2^2 - \omega^2 - i\eta_2\omega) - \alpha_1\alpha_2} E,$$

$$p_2 = \frac{\beta_2(\omega_1^2 - \omega^2 - i\eta_1\omega) + \alpha_2\beta_1}{(\omega_1^2 - \omega^2 - i\eta_1\omega)(\omega_2^2 - \omega^2 - i\eta_2\omega) - \alpha_1\alpha_2} E. \quad (2)$$

Here, $\alpha_1 = (C_{11}l_1/C_{12}l_2)\omega_1^2$, $\alpha_2 = (C_{22}l_2/C_{21}l_1)\omega_2^2$, and $\beta_i = l_i^2/(L_i + L_{i0})$ are the four coefficients ($i$=1, 2); $\omega_i = 1/\sqrt{(L_i + L_{i0})C_{ii}}$ is the plasmon resonance frequency of an isolated gold nanorod; and $\eta_i = \gamma/(1 + L_i L_{i0}^{-1})$ is the damping coefficient (which is slightly smaller than $\gamma$, the collision frequency of free electrons, due to the self-inductance). One can see from Eq. (2) that the asymmetric nanorod dimer has two resonance frequencies. Neglecting the damping of circuits, the (vacuum) resonance wavelength of the dimer satisfies the following equation:

$$2\lambda_\pm^{-2} = \lambda_1^{-2} + \lambda_2^{-2} \pm \sqrt{(\lambda_1^{-2} - \lambda_2^{-2})^2 + 4\kappa^2 \lambda_1^{-2} \lambda_2^{-2}}, \quad (3)$$

where $\lambda_1$ and $\lambda_2$ are the (vacuum) plasmon resonance wavelength of the isolated nanorods. Besides $\lambda_1$ and $\lambda_2$, the ratio between $C_{11}$ and $C_{12}$ ($\kappa \equiv C_{11}/C_{12}$) will also play an important role for the dimer resonance:

$$\kappa = \frac{1}{2}\left(\sqrt{1 + s^2/r^2} - s/r\right). \quad (4)$$

Equations (2), (3) and (4) provide us a basis for studying the plasmonic coupling of gold nanorod dimers in the end-to-end configurations.

With the numerically determined plasmon resonance wavelength of the isolated nanorods [the inset of Fig. 1(a)], we have calculated the dimer resonance using Eq. (3) and (4). For the symmetrical nanorod dimer with $l_1=l_2=120$nm ($\lambda_1 = \lambda_2 = 1023nm$), Eq. (3) predicts a resonance mode at 1149nm, which is very close to the simulated wavelength of 1158nm (the higher energy mode at 931nm is dark due to the dimer symmetry [25]). And for the asymmetrical dimer with $l_1=120$nm and $l_2=150$nm ($\lambda_2 = 1181nm$), the two resonance wavelengths given by Eq. (3) are 978 and 1263nm, in accordance with the numerical simulations 980 and 1275nm. Moreover, good agreement can also be found for the nanorod dimer with $l_1=120$nm and $l_2=180$nm ($\lambda_2 = 1335nm$), where the analytical values around 997 and 1401nm are comparable with the numerical results of 1006 and 1412nm. For comparison, the analytically obtained spectral positions have been indicated in Fig. 1(a) by the vertical arrows.

To look into the coupling effect of nanorod dimer, we calculated the polarizability of nanorods ($\chi_i = p_i / \varepsilon_0 E$) by using Eq. (2). Note that $|\chi_i|$ is responsible for the particle excitation and far-field radiations and $\arg(\chi_i)$ denotes its phase with respect to the incident light. Taking the dimer with $(l_1, l_2)=(120, 180)$ nm as an example, Fig. 2(a) and 2(b) show, respectively, the absolute value and phase of polarizability of the two nanorods. It can be seen from Fig. 2(a) that both nanorods can support the plasmon resonance with the positions locating around 1000 and 1400nm, where the resonance peaks for $|\chi_1|$ and $|\chi_2|$ are present (these peaks are close to the intrinsic resonance of the two nanorods, respectively). Near the wavelength 1000nm, the excitation of short nanorod is obviously stronger than the long nanorod ($|\chi_1|>|\chi_2|$), and $|\chi_2|$ exhibits an asymmetrical lineshape. This unusual lineshape can be understood in terms of the Fano resonance [49]: the resonance of long nanorod benefits from the interference between the nonresonant contribution of incident light and resonant contribution of the short nanorod. And, at the larger wavelength 1400nm, the situation is similar but $|\chi_2|$ is dominant. On the other hand, from 1000nm to 1400nm, $|\chi_2|$ exhibits a large variation while $|\chi_1|$ does not (the short nanorod is almost equally excited at the two resonance wavelengths). These results agree well with the mapped field distribution in Fig. 1(d) and 1(e).

Additional insight can be provided with the phase of polarizability $\phi_i = \arg(\chi_i)$ as shown in Fig. 2(b). At low frequencies ($\lambda > 1500nm$), $\phi_1$ and $\phi_2$ are very small, as the internal electronic motion of both nanorods can keep up with the oscillation of light. When the light frequency approaches the intrinsic resonance frequency of the long nanorod ($\lambda = 1335nm$), a strong plasmon resonance is induced and furthermore it will couple with the short nanorod. In this case, $\phi_2$ together with $\phi_1$ (the latter is driven mainly by the long rod) undergo a $\pi$-phase shift. With the further increase of light frequency (nonresonant wavelength range, $\lambda > 1050nm$), $\phi_2$ varies slightly but $\phi_1$ returns gradually from $\sim \pi$ to very small values (the short nanorod is largely released from the governing of the long one and it oscillates following the incident light). Moreover, when the light frequency is close to the intrinsic frequency of the short nanorod ($\lambda = 1023nm$), the plasmon resonance will be excited and $\pi$-phase shift of $\phi_1$ is resulted. In this case, the long nanorod is mainly driven by the short one and simultaneously, $\phi_2$ also experiences a $\pi$-phase shift with respect to $\phi_1$ (the electronic motion of long nanorod cannot keep up with the driven source, i.e., the short rod). And at high frequencies ($\lambda < 900nm$), both $\phi_1$ and $\phi_2$ are near $\pi$, due to the relatively slow motion of free electrons.

The above results can be highlighted with a plot of phase difference between $\phi_1$ and $\phi_2$ ($\Delta\phi = \phi_2 - \phi_1$, see the solid squares), which shows clearly that, for the long (1400nm) and short (1000nm) wavelength dimer resonance, $\chi_1$ and $\chi_2$ are nearly in-phase and anti-phase, respectively. The charge distribution for the two cases are shown schematically in Fig. 2(a), one corresponds to the attractive lower-energy mode (the upper one) and the other the repulsive higher-energy mode (the lower one). In the former case, the presence of opposite charges near the gap will reduce the potential difference of each nanorod (between the two end faces) and lower the restoring force of free electrons, thus lifting the resonance to longer wavelength (1335nm of the isolated long nanorod versus 1400nm of the dimer). But in the latter case, the potential difference and the restoring force are enhanced due to the presence of charges (near the gap) of the same sign, leading to a blueshift of dimer resonance (1023nm of the isolated short nanorod versus 1000nm of the dimer). It should be noted that, for the symmetrical nanorod dimer, the higher-energy mode with a zero net dipole moment is not optically active [25]. The insets of Fig. 2(a) and 2(b) present, respectively, $|\chi|$ and $\arg(\chi)$ for this case ($l_1=l_2=120$nm), where only the lower-energy mode exists. However, the higher-energy mode can survive when the lateral offset is introduced to the symmetrical dimer [25].

### IV. Cylindrical nanorods of varying gap width

Equations (3) and (4) suggest that the dimer resonance is also dependent on the gap width strongly. To explore this effect, we have studied in this section the plasmon coupling of symmetrical nanorods with the gap width varying. For the cylindrical nanorods having the same size ($l_1=l_2=l$, $\omega_1 = \omega_2 = \omega_0$, and $\eta_1 = \eta_2 = \eta_0$), the dipole moment of each nanorod can be simplified using Eq. (2) as

$$p = \frac{\beta_0}{(1-\kappa)\omega_0^2 - \omega^2 - i\eta_0\omega}. \tag{5}$$

Correspondingly, the (vacuum) dimer resonance wavelength is simply

$$\lambda = \lambda_0 / \sqrt{1-\kappa}, \tag{6a}$$

and the fractional wavelength shift of the dimer becomes

$$\frac{\Delta\lambda}{\lambda_0} = \frac{1}{\sqrt{1-\kappa}} - 1. \tag{6b}$$

Here, $\lambda_0$ is the (vacuum) plasmon resonance wavelength of the isolated nanorod, $\Delta\lambda = \lambda - \lambda_0$ is the wavelength shift, and $\kappa$ is determined by Eq. (4). Our result presents a new Plasmon Ruler Equation for the nanorod dimer, which suggests that the fractional wavelength shift $\Delta\lambda/\lambda_0$ is uniquely determined by the parameter $\kappa$. According to Eq. (4), it means $\Delta\lambda/\lambda_0$ should be related to gap width scaled for the rod radius (s/r) rather than for the rod length (s/*l*) as taken previously [25, 38, 39]. This point can be understood as follows: At close approaches, the coupling of nanorods is mainly mediated by the terminal surface charges near the rod gap, where the rod radius and gap width are two determining geometrical parameters. More explicitly, due to the coupling, the related potential modification of nanorods is dependent on $C_{11}$ and $C_{12}$, which are functions of rod radius and gap width (also see Sec. III). Another remarkable feature of Eq. (6b) is that no fitting parameters are used, thus providing great convenience for the study. In contrast, the widely used exponential size dependence [Eq. (1)] relies on two fitting coefficients, which were found to vary with the dimer size significantly [38, 40].

To test the validity of Eq. (6), we have calculated the far-field scattering spectrum of nanorod dimer with the FDTD method and the results are presented in Fig. 3(a)-3(f). The dimer resonance corresponds to the maximum of absolute value of dipole moment [see Eq. (5)] and the peak of scattering spectrum. Here, the rod length is *l*=150nm, the rod radius is *r*=15nm, and the gap width is decreased from (a) 75nm to (b) 45nm, (c) 30nm, (d) 15nm, (e) 6nm, and (f) 3nm. One can see that the decrease of gap width gives rise to a significant redshift of the peak of scattering spectrum or dimer resonance wavelength. For the gap width s=3nm, for example, the dimer resonance locates at 1582nm; compared with the isolated nanorod ($\lambda_0 = 1181nm$), a wavelength shift of 400nm or fractional shift of 34% is obtained. In addition, the resonance wavelength of nanorod dimer has also been determined analytically with Eq. (6). Figure 3(g) gives a comparison between the numerical (the solid squares) and analytical (the solid line) calculations, which shows a good agreement. In spite of the fitting parameters, the accuracy of exponential size scaling has been proven to be rather limited for the small particle separations (s/*l*<0.09 or s<13.5nm here) [25]. In contrast, the agreement achieved here is still good even when the particle separation is down to 3nm (or s/*l*=0.02). However, we have to point out that, here, when the gap width is smaller than 1nm, the agreement between theory and simulations will get worse (not shown here). In the extremely small particle separations, the quantum effect will become very important, where the free electrons may tunnel from one nanorod to the other. In this case, the optical properties of the nanorod dimer may approach that of a single nanorod with a double rod length. Such a particle separation (~1nm) may set a limit of validity for our theoretical results.

The dimer resonance will be accompanied by a strong field enhancement in the dielectric gap. Here the near-field distribution (the axial E-field) in the middle plane of the gap has been simulated and mapped as shown in the inset of Fig. 3(a)-3(f). When the gap width *s* is larger than 40nm [see (a) and (b)], the field is rather "dark". When s is decreased to 30 or 15nm [see (c) and (d)], the field becomes "bright". And when s is down to 6 or 3nm [see (e) and (f)], the field gets very "hot". The inset of Fig. 3(g) gives the E-field enhancement factor as a function of gap width, where an increase of electric field with decreasing gap width is clearly manifested. It is worthy of noticing that there are two reasons responsible for this effect. On the one hand, with the decrease of gap width, the detection spot (the gap center) will be closer to the end faces and "feels" a larger electric field. On the other hand, according to Eq. (5), the maximal absolute value of dipole moment $|p|_m$ is inversely proportional to the dimer resonance frequency. The decrease of gap width will reduce the resonance frequency and increase the dipole moment $|p|_m$, thus resulting in an amplification of the near field. For the gap width s=3nm, the electric field (or intensity) in the gap center is about 186 (or 35000) times larger than that of the incident light. Such huge field amplification can be employed to enhance the light-matter interactions [1].

Since the fractional wavelength shift $\Delta\lambda/\lambda_0$ is uniquely determined by the gap width scaled for the rod radius *s/r*, the distance dependence of rod dimers of different sizes may be mapped with one line. With the use of Eq. (6b) and Eq. (4), $\Delta\lambda/\lambda_0$ as a function of $s/r$ has been plotted and the result is shown as the solid line in Fig. 4(a). Besides the numerical data presented in Fig. 3(g) [or triangles replotted in Fig. 4(a)], we have performed additional numerical calculations to seek more evidence for our results. Figure 4(c) shows the scattering spectra of a nanorod dimer with the varying gap width, where the rod length is $l$=90nm, the rod radius is r=15nm, and the gap width is s=30, 20, 15, 10, 5, 3, 1nm, respectively (the spectrum of an isolated nanorod is denoted by the arrow). The dimer resonance wavelengths extracted from the spectra have been processed and given in Fig. 4(a) by the solid circles. It can be seen that the extracted data match well the theoretical solid line (note that a deviation is developed for the gap width s=1nm). In addition, the scattering spectra of another nanorod dimer have also been simulated [see Fig. 4(d)], where $l$=100nm, r=10nm, and s=50, 25, 10, 5, 2nm, respectively. The results are shown in Fig. 4(a) by the solid squares, suggesting again the good agreement with the solid line.

In previous studies, the fractional wavelength shift of the nanorod dimer has been mapped as a function of gap width scaled by the rod length $s/l$ [25, 38, 39]. For comparison, here in Fig. 4(b), $\Delta\lambda/\lambda_0$ versus $s/l$ for three groups of nanorod dimers have also been plotted respectively, where the solid lines represent our analytical calculations and the symbols the numerical results. It is found that the dimers with ($l$=100nm, r=10nm) and ($l$=150nm, r=15nm), corresponding to the same aspect ratio, share a common line. However, the nanorod dimers with different aspect ratios will have their respective $s/l$ dependence. For example, the line with the solid circles ($l$=90nm, r=15nm) is displaced significantly from that with the solid squares ($l$=100nm, r=10nm). It can be concluded that the gap width in units of the rod length is inadequate for determination of the fractional wavelength shift.

## V. Rectangular nanorods of varying gap width

Cylindrical nanorods can be made with the chemical synthesis, which yields single crystals of well-defined shape and high quality [25]. However, such a method is followed by the challenge in arranging the nanorods regularly. The modern nanofabrication technology such as electron beam lithography (EBL) or focused-ion beam (FIB) milling may provide alternative methods for the fabrication of gold nanorods or nanorod dimers [33, 34, 39]. In this case, the rectangular rather than cylindrical nanorod dimers can be obtained. Then one may ask, what about the resonance feature of rectangular nanorod dimers? A simple investigation of this case will be given in this section.

Although the rectangular nanorod dimer will make a difference from the cylindrical one, our study suggests that the previous results still hold except that $\kappa$, the ratio between $C_{11}$ and $C_{12}$, needs a modification. For the rectangular nanorod dimer with a gap width $s$ and cross-sectional area $a \times b$, we have

$$\kappa = \frac{a\ln\dfrac{\tau+b}{\tau-b} + b\ln\dfrac{\tau+a}{\tau-a} - 4s\cdot\tan^{-1}\dfrac{ab}{2s\tau}}{4a\ln\dfrac{\sigma+b}{a} + 4b\ln\dfrac{\sigma+a}{b}}. \quad (7)$$

Here, the coefficients $\sigma = \sqrt{a^2+b^2}$ and $\tau = \sqrt{a^2+b^2+4s^2}$. Equation (7) shows a rather complicated relationship between $\kappa$ and the geometrical parameters. However, the calculations suggest that a significant simplification can be made.

Using the Eq. (6b) and (7), the dependence of fractional wavelength shift of symmetrical rectangular nanorod dimers on the gap width has been calculated. Here we considered two cases: (i) the nanorod has a square cross-section with $a=b=28.3nm$; (ii) the nanorod has a rectangular cross-section with a=20nm and b=40nm. Note that, in both cases, the nanorods have the same cross-sectional area. The calculated results are shown in Fig. 5(a) by the solid (i) and dash (ii) lines respectively, where the gap width has been normalized to the effective radius $r_{eff}$ of rectangular nanorods ($r_{eff} = \sqrt{ab/\pi} \approx 16nm$). One can see that the theoretical line-shape for the two cases almost coincides with each other. The inset of Fig. 5(a) gives a detail amplification of the line-shape around $s/r_{eff} =1.08$, suggesting that the difference of $\Delta\lambda/\lambda_0$ between the two lines is less than 0.2%. Therefore, the fractional wavelength shift of rectangular nanorod dimers is not sensitive to the transverse nanorod sizes once the cross-sectional area is fixed. Such a conclusion was also supported by the additional analytical calculations.

To confirm the above results, the scattering spectra of rectangular nanorod dimers with the varying gap width have been simulated. In case (i), the gap width is set as s=80, 40, 20, 10, and 4nm, respectively ($a=b=28.3$nm); and in case (ii), the gap width is s=60, 30, 15, 6, and 3nm, respectively (a=20nm, b=40nm). The length of nanorods is fixed as $l$=120nm. The simulation results for the dimers are presented in Fig. 5(b) and 5(c), respectively (the arrow denotes the spectrum of an isolated nanorod). Similar to the cylindrical nanorod dimers studied above, a decrease of gap width causes a distinct redshift of rectangular dimer's resonance. The numerical results extracted from the scattering spectra have been mapped in Fig. 5(a) by the symbols, which match well the theoretical lines.

In addition, we have also compared the distance dependence of rectangular nanorod dimer with that of a cylindrical one. Figure 5(a) and its inset plotted the corresponding line-shape for the cylindrical nanorod dimer (the dot line), where the rod radius is $r = r_{eff} = 16nm$ [here Eq. (6) and (4) are employed]. Remarkably, the resulted line-shape also shows a good accordance with that of rectangular dimers. This indicates that the fractional wavelength shift of rectangular nanorod dimer can be approximated by using an equivalent cylindrical dimer which owns the same cross-sectional area. The conclusion, also supported by the calculations with other parameters, provides a great simplification for the study of rectangular nanorod dimers.

## VI. Wave-guiding in a chain of nanorods

In previous sections, the study has been limited to the plasmon coupling within a pair of nanorod particles. When the metallic particles are aligned in a linear chain, the energy transfer along the chain may be mediated by the strong near-field coupling. Such a linear chain consisting of metallic particles has been used as the subwavelength plasmonic waveguides [6, 7], which may overcome the diffraction limit encountered conventionally. Up to date, most of researches have employed the spherical nanoparticles as the building blocks [6, 50-55]. This enables the point dipole model to be used, thus providing convenience for the theoretical analysis. Nonetheless, the plasmon resonance of the nanosphere owns a rather poor tuning characteristic, which sets a barrier for the waveguides to work at the desired wavelength. Moreover, the curved surface of spherical particles is not optimal for confining the fields to the particle junctions.

Here we employ the gold nanorods as the building blocks of the plasmonic waveguides. The plasmon resonance of nanorods (and thus the working frequency band of waveguides) can be tuned conveniently from the visible to near infrared by changing the aspect ratio [48, 56, 57]. We consider the case that the gold nanorods are arranged in a linear chain with the end-to-end configurations, thus the longitudinal excitation of chain is supported and the fields are strongly confined in the nanosized dielectric gaps. Theoretically, however, the gold nanorods cannot be treated as the point dipoles, especially when the gap width of nanorods is smaller than half the rod length. Therefore, at close approaches of nanorods, the point dipole model, which was utilized previously for the spherical nanoparticle waveguides [50-54], will not be applicable. In the following, nonetheless, the coupled LC-resonator scheme employed above may provide a solution for the question.

Assume that the cylindrical nanorods with the length $l$ and radius $r$ are aligned in a linear array with the gap width $s$ and lattice period $d$ ($d=l+s$). To investigate the dispersion relation of this waveguide, we calculate the electric potential at the end faces and apply the LC-circuit equation to the nanorods. Considering that the longitudinal excitation can induce the positive and negative charges in each nanorod, the electric potential at the end faces actually benefits from all the charges (with the respective retarded potential). Nonetheless, a fundamental approximation can be made by only counting the nearest-neighbor interactions. In this case, the electric potential difference of the $n$th nanorod (between the centers of two end faces) will be $\Delta u_n = q_n / C_{11} - (q_{n+1} + q_{n-1}) / C_{12}$, where $q_n$ is the charge carried by the $n$th nanorod ($C_{11}$ and $C_{12}$ are defined as before). The calculations with the LC-circuit equation and propagating wave solutions yield the following waveguide dispersion:

$$\omega^2/\omega_0^2 = 1 - 2\kappa\cos(k_1 d)\cosh(k_2 d),$$
$$\eta_0\omega/\omega_0^2 = 2\kappa\sin(k_1 d)\sinh(k_2 d). \quad (8)$$

Here, $\omega_0$ is the resonance frequency of an isolated nanorod, $\eta_0$ is the damping coefficient, and $\kappa$ is the ratio between $C_{11}$ and $C_{12}$ [see Eq. (4)]; $k_1$ and $k_2$ is the real and imaginary part of the propagation constant, respectively.

With the use of Eq. (8), we have calculated the dispersion relation and normalized energy decay length ($N = 1/2k_2 d$) for an infinite chain of gold nanorod particles. Here, the rod length is $l$=120nm, the rod radius is r=15nm, and the gap width is s=10nm. The results are presented in Fig. 6(a) by the solid lines. The working frequency band of waveguide extends from $0.63\omega_0$ to $1.26\omega_0$ (the band edge is defined using $N$=1), corresponding to a relative band width about 68%. At the frequency of $\omega_m = 0.92\omega_0$ (or central wavelength $\lambda = 1112nm$), a maximal energy decay length of $N$=6.8 is predicted; this corresponds to a group velocity of longitudinal plasmon wave of $0.22c$ or $6.7\times10^7 m/s$. For comparison, we have also simulated the wave propagation behavior in such a waveguide employing a finite 40-rod chain. The obtained energy decay length as a function of normalized frequency is indicated in Fig. 6(a) by the solid circles. A reasonable agreement between the analytical and numerical calculations can be seen. However, the maximal decay length obtained numerically shows a slight redshift ($\omega_m = 0.90\omega_0$ or $\lambda = 1140nm$) and is increased ($N$=7.4) compared with the theoretical predictions. This deviation can be attributed to the finite chain-length of simulation as well as the nearest-neighbor approximation of the model, as, near the resonance ($\omega \sim \omega_0$), the nanorod particles are strongly excited and the far-neighbor interactions also make a contribution. In addition, the typical dependence of electric field on the propagation length L, which was calculated numerically for the excitation frequency $\omega_m = 0.90\omega_0$, is shown in Fig.

6(b). One can see that the fields, which are strongly confined in the dielectric gaps and is one order of magnitude larger than that in nanorods, decay exponentially with the length (the oscillation for L>3000nm is due to the reflection of plasmon wave at the waveguide terminal). The energy decay length $N$=7.4 (or field decay length about 1920nm) can be deduced from the inset of Fig. 6(b), corresponding to the signal attenuation coefficient of $2k_2 = 1.04 \times 10^6 m^{-1}$ or $4.5 dB/\mu m$. In contrast, the commonly studied particle plasmon waveguides have employed the metallic nanospheres with the diameter 50nm and spacing 75nm (the central wavelength is ~350nm), which yields a relative band width of ~27%, group velocity of 0.19$c$, and field decay length of 1500nm (or signal attenuation $6 dB/\mu m$) [50, 52].

    The bandwidth and central working frequency of the waveguide can be modulated by varying the gap width of nanorods (the rod sizes are fixed). With the decrease of gap width, the upper and lower band edges are blueshifted and redshifted respectively, thus enlarging the pass band [see Fig. 6(c)]. Simultaneously, the central working frequency $\omega_m$ (for achieving the maximal decay length) shifts to the longer wavelength and the maximal energy decay length is increased [see Fig. 6(d)]. For the gap width s=3nm, for instance, the relative band width is about 120%. Moreover, the central wavelength is redshifted to 1430nm, and the normalized energy decay length $N$ attains 11.0 (corresponding to a field decay length ~2700nm). The simulation results are presented in Fig. 6(c) and 6(d) by the solid circles or squares, in accordance with the analytical calculations (the solid lines). Furthermore, the field (intensity) distributions simulated for various gap widths and central working wavelengths are shown in Fig. 6(e), 6(f), and 6(g), respectively. Besides the increase of decay length with the decreasing gap width, one can see that the fields are well localized in the waveguide. For the gap width s=3nm, the lateral mode size determined by the 1/$e$ energy decay is about 50nm, which is only 3.5% of the working wavelength 1430nm (here a mode size of $\sim \lambda/28$ compares with $\sim \lambda/5$ of the spherical particle waveguides [50, 52]). This deep subwavelength mode size is desirable for the future photonic or plasmonic integrations [58].

## VII. Conclusions

The plasmon coupling of gold nanorod particles at close approaches has been studied. We have considered several cases, including the nanorod dimer with the asymmetrical rod lengths, symmetrical rod dimer with the varying gap width, and waveguiding in a linear chain of nanorod particles. Our findings can be summarized as follows. First, the coupling between the asymmetrical nanorods of a dimer results in two plasmon modes: one is the attractive lower-energy mode and the other the repulsive high-energy mode. Using the coupled LC-resonator scheme, the spectral position of dimer resonance can be determined analytically. Second, the fractional wavelength shift of a symmetrical cylindrical nanorod dimer with the varying gap width is uniquely dependent on the gap width scaled for the rod radius ($s/r$) rather than gap width in units of the rod length ($s/l$). A new Pasmon Ruler Equation for the nanorod dimer has been given, where no fitting parameters are used. Third, the plasmon coupling of symmetrical rectangular nanorod dimer (with the varying gap width) is almost independent of the transverse sizes once the cross-sectional area of nanorod is fixed. Using the effective rod radius, the coupling response of rectangular rod dimer can be well approximated with that of the cylindrical one. And fourth, a linear chain of gold nanorod particles can be employed for the efficient plasmonic wave-guiding. A field decay length up to 2700nm with the lateral mode size about 50nm ($\sim \lambda/28$) has been suggested. These results may provide new opportunity for the understanding and application of the plasmon coupling effect. Although our interest has been focused on the end-to-end configurations, the method can also be extended to study other cases, such as the nanorod dimers arranged side-to-side [25].

**Acknowledgment.** This work was supported by the National Natural Science Foundation of China (Grant No. 10804051 and 10874079), by the State Key Program for Basic Research of China (Grant No. 2010CB630703).

**Figure captions**

**Figure 1**: (a) Far-field scattering of cylindrical gold nanorod dimer, where the black open circles, the green open squares, and the red solid circles correspond, respectively, to the rod length ($l_1$, $l_2$)=(120, 120), (120, 150), and (120, 180) nm. The analytically calculated peak positions are indicated by the vertical arrows. The inset shows, from the left to right, the spectra of an isolated nanorod with the length 120, 150, and 180nm, respectively. The schematic view of nanorod dimer is shown in (b). (c) The axial E-field distribution of dimer with ($l_1$, $l_2$)=(120, 120) nm (at the resonance wavelength 1162nm). (d) and (e) The axial E-field distribution associated with the short (1006nm) and long (1412nm) wavelength resonance of dimer with ($l_1$, $l_2$)=(120, 180) nm. Here, the rod radius and gap width are fixed as r=s=15nm.

**Figure 2**: (a) Absolute value of the polarizability for the short (the open circles) and long (the solid circles) nanorods of an asymmetrical dimer with ($l_1$, $l_2$)=(120, 180) nm. The charge distributions for two plasmon resonances are shown schematically. (b) The corresponding phases of the polarizability of the nanorods. The solid squares represent the phase difference between two polarizabilities. As a comparison, $|\chi|$ and $arg(\chi)$ for a symmetrical nanorod dimer with ($l_1$, $l_2$)=(120, 120) nm are plotted in the insets of (a) and (b), respectively. Here, r=s=15nm.

**Figure 3**: Far-field scattering and near-field (the axial E-field in the middle plane of the gap) distribution of a symmetrical cylindrical nanorod dimer with the varying gap width s: (a) 75nm, (b) 45nm, (c) 30nm, (d) 15nm, (e) 6nm, and (f) 3nm. The dependence of plasmon resonance wavelength on the gap width is shown in (g), where the line and squares represent, respectively, the analytical and numerical results (the inset gives the E-field enhancement factor as a function of gap width). Here, the rod length is $l$=150nm and the rod radius is $r$=15nm.

**Figure 4**: Relative wavelength shift of longitudinal plasmon resonance as a function

of (a) gap width scaled for rod radius *s/r* and (b) gap width scaled for rod length *s/l*, where the solid lines and the symbols represent, respectively, the analytical and numerical calculations. Far-field scattering of a symmetrical cylindrical nanorod dimer with the varying gap width s are shown in (c) for l=90nm, r=15nm, and s=30, 20, 15, 10, 5, 3, 1nm, and in (d) for l=100nm, r=10nm, and s=50, 25, 10, 5, 2nm. The spectrum of an isolated nanorod is denoted by the arrow.

**Figure 5**: (a) Fractional wavelength shift of symmetrical rectangular nanorod dimer as a function of gap width scaled for the effective rod radius *s/r$_{eff}$*, where the lines and symbols represent, respectively, the analytical and numerical calculations (the details of line-shape around *s/r$_{eff}$* =1.08 are depicted in the inset). The far-field scattering spectra of rectangular nanorod dimers with the varying gap width s are shown in (b) *a=b*=28.3nm, s=80, 40, 20, 10, 4nm, and (c) a=20nm, b=40nm, s=60, 30, 15, 6, 3nm. The spectrum of an isolated nanorod is denoted by the arrow. Here, the length of nanorod is fixed as *l*=120nm.

**Figure 6**: Wave-guiding in a linear chain of gold nanorod particles. (a) Analytically determined dispersion relation and normalized energy decay length *N* for an infinite chain with gap width s=10nm. The circles represent the decay length *N* obtained with FDTD simulations (for a finite 40-rod chain). (b) Simulated decay behavior of electric field with the propagation length *L*, where the excitation frequency corresponds to $\omega/\omega_0 = 0.90$ in (a). The inset shows ln(I/I$_0$) in the gap center as a function of *L/d* (the circles) and the line denotes a linear fit of data. (c) Bandwidth and (d) maximal energy decay length and the corresponding excitation frequency vs. gap width (the lines and symbols represent the analytical and numerical results, respectively). (e), (f), and (g) present the field intensity distributions for various gap width and central working wavelength. Here, the rod sizes are fixed as l=120nm and r=15nm.

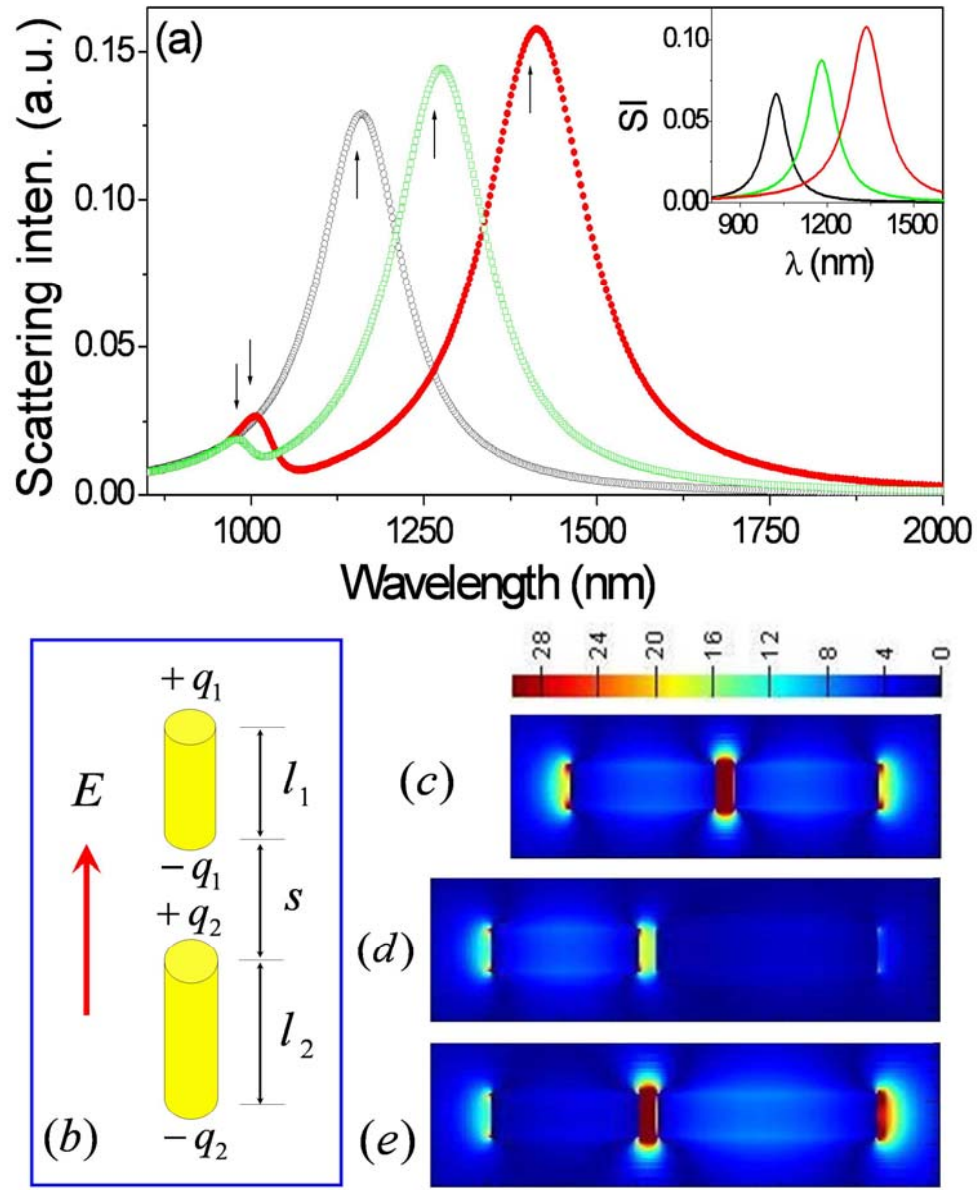

**Figure 1**

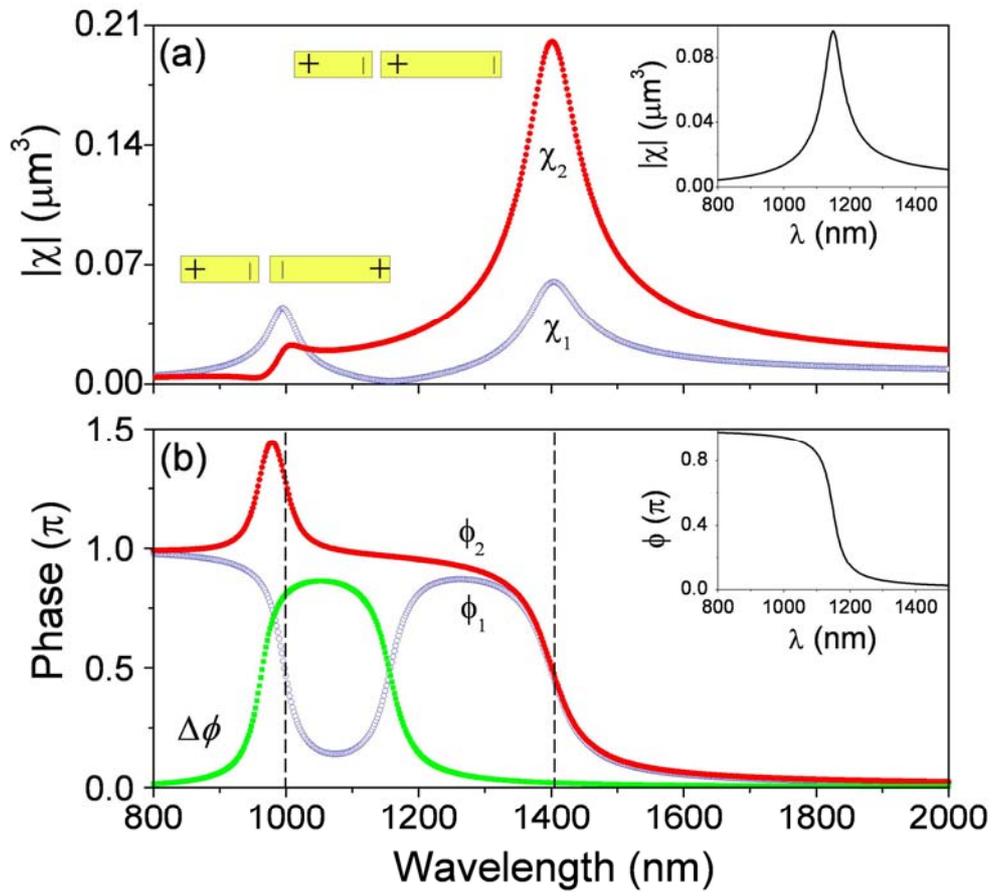

**Figure 2**

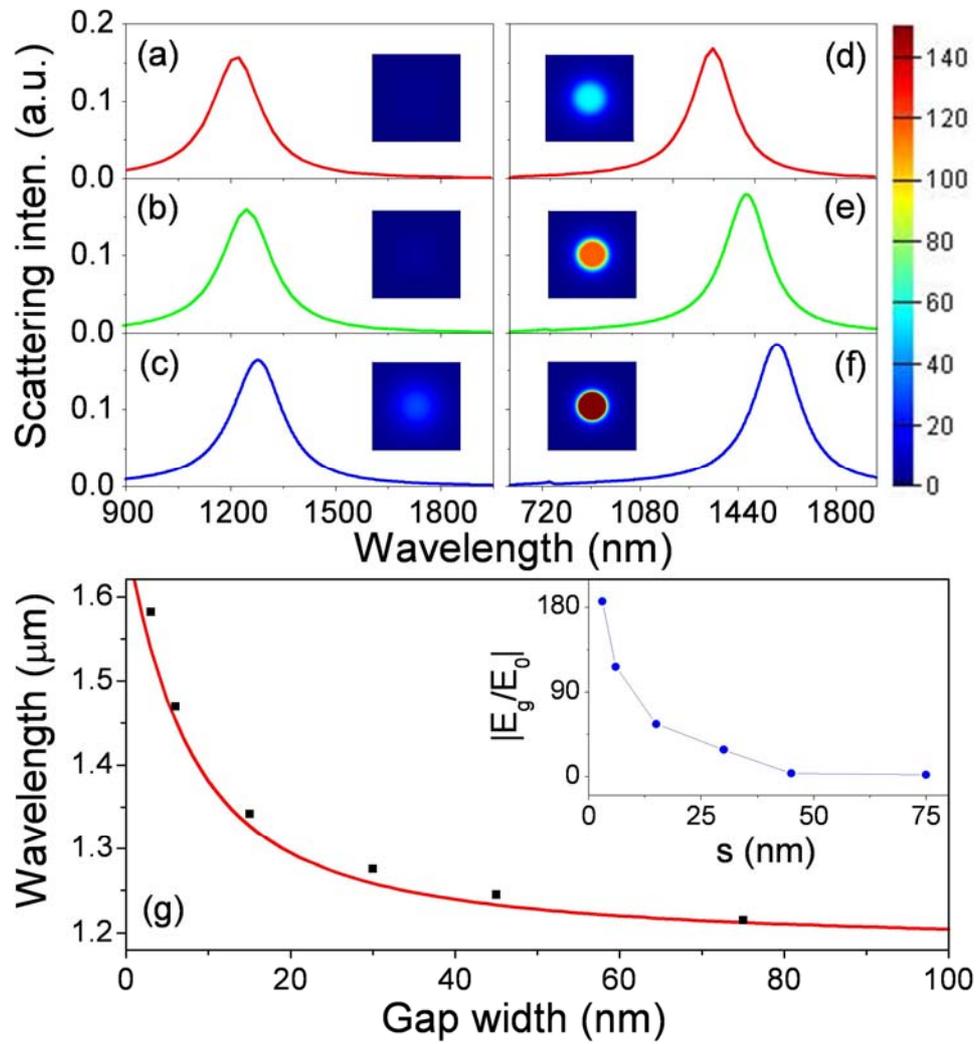

**Figure 3**

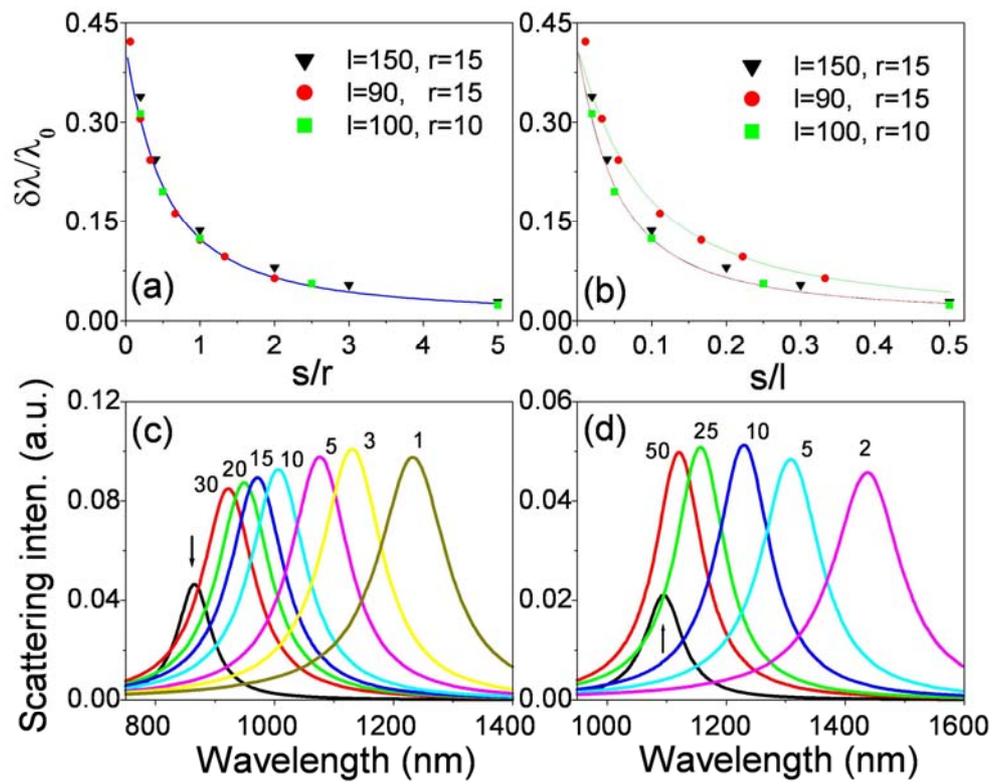

**Figure 4**

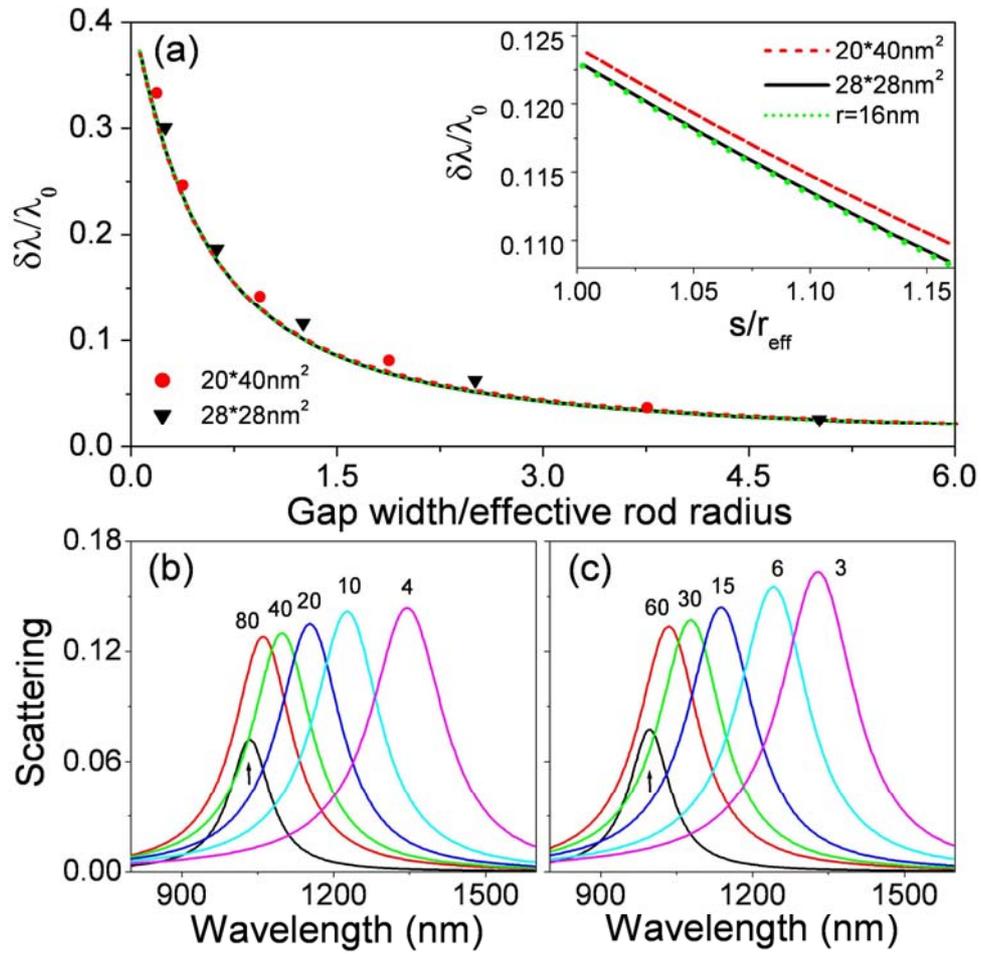

**Figure 5**

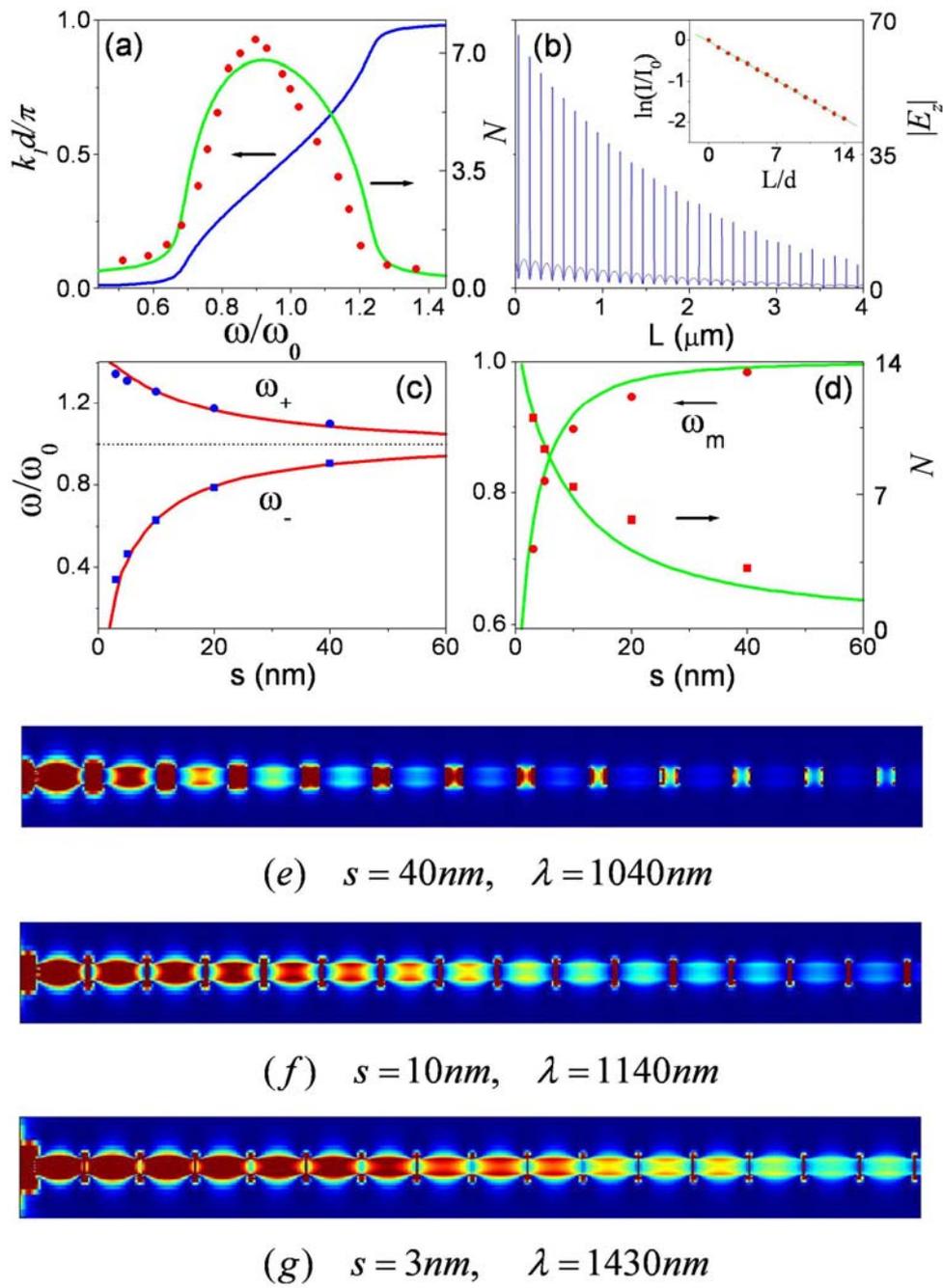

**Figure 6**